\documentclass[twoside,12pt]{article}
\usepackage{indentfirst}
\usepackage{bm}
\usepackage{epsfig}
\usepackage{graphicx}
\usepackage{dcolumn}
\usepackage{bm}
\usepackage{xcolor}
\usepackage{url}
\usepackage{makecell}
\usepackage{subfigure}
\usepackage{multirow}
\usepackage{booktabs}
\usepackage{verbatim}
\usepackage{diagbox}
\usepackage[marginal]{footmisc}
\usepackage{amssymb}

\begin{document}


\begin{center}
\large\bf Valence Quark Distributions in Pions: Insights from Tsallis Entropy
\end{center}

\begin{center}
\rm Jingxuan Chen$^{\rm a,b,c,d)*}$, \ \ Xiaopeng Wang$^{\rm c,d,e)*}$, \ \ Yanbing Cai$^{\rm f,g)\dag}$,\ \ Xurong Chen$^{\rm c,d)\ddagger}$,\ and  \ Qian Wang$^{\rm a,b,g)\xi}$
\end{center}

\footnotetext{\hspace*{-.05cm}\footnotesize $^*$These authors contributed equally: Jingxuan Chen and Xiaopeng Wang.}
\footnotetext{\hspace*{-.05cm}\footnotesize $^\dag$E-mail: yanbingcai@mail.gufe.edu.cn}
\footnotetext{\hspace*{-.05cm}\footnotesize $^{\ddagger}$E-mail: xchen@impcas.ac.cn}
\footnotetext{\hspace*{-.05cm}\footnotesize $^\xi$E-mail: qianwang@m.scnu.edu.cn}

\begin{center}
\begin{footnotesize} \sl
${}^{\rm a)}$ Key Laboratory of Atomic and Subatomic Structure and Quantum Control (MOE), Guangdong Basic Research Center of Excellence for Structure and Fundamental Interactions of Matter, Institute of Quantum Matter, South China Normal University, Guangzhou 510006, China\\
${}^{\rm b)}$ Guangdong-Hong Kong Joint Laboratory of Quantum Matter, Guangdong Provincial Key Laboratory of Nuclear Science, Southern Nuclear Science Computing Center, South China Normal University, Guangzhou 510006, China\\
${}^{\rm c)}$ Institute of Modern Physics, Chinese Academy of Sciences, Lanzhou 730000, China\\
${}^{\rm d)}$ School of Nuclear Science and Technology, University of Chinese Academy of Sciences, Beijing 100049, China\\
${}^{\rm e)}$ School of Nuclear Science and Technology, Lanzhou University, Lanzhou 730000, China\\
${}^{\rm f)}$ Guizhou Key Laboratory in Physics and Related Areas, Guizhou University of Finance and Economics, Guiyang 550025, China\\
${}^{\rm g)}$ Southern Center for Nuclear-Science Theory (SCNT), Institute of Modern Physics, Chinese Academy of Sciences, Huizhou 516000, China\\
\end{footnotesize}
\end{center}

\begin{center}
\begin{minipage}{15.5cm}
\parindent 20pt\footnotesize
We investigate the valence quark distributions of pions at a low initial scale ($Q^2_0$) by employing Tsallis entropy, a non-extensive measure that effectively captures long-range correlations among internal constituents. Utilizing the maximum entropy approach, we adopt two distinct functional forms and fit experimental data through the elegant GLR-MQ-ZRS evolution equation to derive the model parameters. Our findings indicate that the resulting valence quark distributions provide an optimal fit to experimental data, with the values of the $q$ parameter deviating from unity. This deviation indicates the significant role that correlations among valence quarks play in shaping our understanding of pion internal structure. Additionally, our computations of the first three moments of pion quark distributions at $ Q^2 = 4 \, \mathrm{GeV}^2$ display consistency with other  theoretical models, thereby reinforcing the importance of incorporating valence quark correlations within this analytical framework.
\end{minipage}
\end{center}

\begin{center}
\begin{minipage}{15.5cm}
\begin{minipage}[t]{2.3cm}{\bf Keywords:}\end{minipage}
\begin{minipage}[t]{13.1cm}
Tsallis entropy, maximum entropy method, valence quark distributions
\end{minipage}\par\vglue8pt

\end{minipage}
\end{center}

\section{Introduction}
\label{sec:intro}
Partons distribution functions (PDFs) are key ingredients in quantum chromodynamics (QCD) collinear factorization theorem, in which the scattering processes can be factorizing into the long-distance physics and the short-distance hard scattering cross sections. PDFs defined as the probability density for finding a parton with longitudinal momentum fraction $x$ at resolution scale $Q^2$ are the part of the long-distance physics. In contrast to the hard scattering cross sections which can be calculated in perturbative QCD, PDFs have non-perturbative property. Therefore, the determination of the PDFs of the hadron has been a challenge to theory study. In recent years, the investigation of PDFs for the pion meson has been of great interest, for instance the latest calculation in Ref.~\cite{Fan:2021} and references therein. The pion, consistent of two valence quarks, is the lightest hardon, which is the Nambu-Goldstone boson of dynamical chiral symmetry breaking ~\cite{Goldstone:1961,Nambu:1961}. In addition, the pion is also the most important mediated particle of interaction between hadrons, for instance it is the mediated particle of nuclear force proposed in Ref.~\cite{Yukawa:1935}. As the result, the study of the internal structure of pion is useful to understand both dynamical chiral symmetry breaking and non-perturbative QCD.

The $x$-dependence of parton density functions at low resolution scale are usually described by parameterizations of the non-perturbative input and the parameters in it are obtained by fitting observables to experimental data. However, compared to nucleon PDFs, our knowledge of pion PDFs is limited due to the scarcity of experimental data. Fortunately, there are many theoretical research on pion PDFs focuses on determining the quark distribution, such as chiral-quark model ~\cite{Nam:2012vm,Watanabe:2016lto,Watanabe:2017pvl,Nematollahi:2018iac} , constituent quark model ~\cite{Suzuki:1997wv,Szczepaniak:1993uq,Frederico:1994dx}, basis light-front quantization (BLFQ)~\cite{Lan:2019vui,Lan:2019rba}, light-front holographic QCD (LFHQCD)~\cite{deTeramond:2018ecg}, QCD sum rule~\cite{Ioffe:1999hz}, Dyson-Schwinger equation (DSE)~\cite{Ding:2019qlr,Cui:2020tdf,Shi:2018mcb,Bednar:2018mtf}, lattice QCD (LQCD)~\cite{Gao:2020ito} and maximum entropy method (MEM) ~\cite{Han:2018wsw,Zhang:2023oja}. The gluon distribution of the pion has been investigated using holographic QCD~\cite{Watanabe:2019zny}, DSE~\cite{Ding:2019qlr,Cui:2020tdf,Freese:2021zne,Chang:2021utv} and LQCD~\cite{Fan:2021}. Notably, Ref.~\cite{Chang:2021utv}  presents a comparison between LQCD and DSE. These two methods demonstrate good agreement, providing valuable insights into the origins of hadron masses. Furthermore, the pion structure function or PDFs is also predicted by NJL model~\cite{Bentz:1999gx,Shigetani:1993dx,Davidson:1994uv,Shakin:1996fa,Hutauruk:2016sug,Davidson:2001cc}.  In addition, there are also some data-driven approaches to investigate the pion distribution functions (see Ref.~\cite{Lu:2023yna} and references therein). Collectively these efforts enhance our comprehension of the internal structure of pions.

Among the various models, MEM offers a straightforward yet efficient approach for determining pion valence quark distributions~\cite{Han:2018wsw}. It is proposed as an alternative to conventional global fits for extracting parton distribution functions (PDFs)~\cite{Wang:2014lua}. In the realm of information theory, entropy serves as a quantitative measure of uncertainty and has gained increasing significance within the context of physical theories. It is essential to emphasize that parton distributions are not direct experimental observables; rather, they represent probabilistic constructs. Consequently, these distributions are fundamentally linked to the principles of information entropy. The relationship between entropy and gluon distribution has been notably explored by Kharzeev et al.~\cite{Kharzeev:2017}. Furthermore, the maximum entropy principle, as articulated by Jaynes~\cite{Jaynes:1957}, plays a pivotal role in establishing probability distributions that minimally incorporate external biases. The core premise of MEM is to derive the most unbiased distribution based on available information. In the studies concerning parton distributions, MEM proves particularly advantageous in scenarios where information derived from Quantum Chromodynamics (QCD) calculations is limited~\cite{Wang:2014lua}.

In MEM, the known properties of the hadron serve as testable information, while the valence quark distributions are determined by maximizing entropy. In Ref.\cite{Han:2018wsw}, the method of Maximum Entropy (MEM) is first employed to model the non-perturbative input associated with pions. At a very low resolution probe, valence constituents dominate in the quark-parton model. Based on the Von Neumann entropy, which satisfies additivity properity, the entropy of valence quark distributions at initial scale $Q^2_0$ is defined as
\begin{equation}
S=-\int_0^1[u_\nu(x,Q^2)\mathrm{ln}(u_\nu(x,Q^2))+\bar{d}_\nu(x,Q^2)\mathrm{ln}(\bar{d}_\nu(x,Q^2))]dx.
\label{eq:S_Von_Neumann}
\end{equation}
With this entropy, the valence quark distribution is well described when the momentum contributed from gluons is nonzero. However, Eq.~(\ref{eq:S_Von_Neumann}) is obtained under the ansatz that the interaction between valence quarks is negligible. It is more reasonable to describe the internal structure of pion in terms of non-extensive entropy, because the valence quarks are correlated. Tsallis entropy, which is a kind of non-extensive entropy~\cite{Tsallis:1987eu,Tsallis:2011rbr,Tsallis:2017fhh}, is defined as
\begin{equation}
\label{eq:tsallis_entropy}
S_q(p)=\frac{1}{q-1}(1-\int(p(x))^q dx) , ~(q\neq1),
\end{equation}
where $p(x)$ is probability distribution function, and the index $q$ represents the correlation strength of a system. When $q\rightarrow 1$, Eq.~(\ref{eq:tsallis_entropy}) is reduced to Von Neumann entropy
\begin{equation}
    \label{eq:shaonn}
    S=-\int p(x)\ln p(x) \mathrm{d}x.
\end{equation}
Tsallis entropy satisfies pseudo-additivity, which is represented as
\begin{equation}
    \label{eq:pseudo_additivity}
    S_q=S_q(u_\nu)+S_q(\bar{d}_{\nu})+(1-q)S_q(u_\nu)S_q(\bar{d}_\nu).
\end{equation}
Tsallis entropy is one-parameter extension of the Von Neumann entropy, it has more applicable and flexible for the description of a system with long-range correlation. In this study, we shall employ the Tsallis entropy to determine the valence quark distributions of pion based on the maximum entropy method. Our results indicate that the correlation between valence quarks plays an important role in describing the valence quark distributions of pion.

The arrangement of this article is as follows. In Sec.~2, we briefly introduce a naive non-perturbative input for valence quarks of pion and the definition of MEM with Tsallis entropy. In Sec.~3, based on the simplified version of the modified DGLAP evolution with GLR-MQ-ZRS corrections, the results from MEM are compared to the experimental data and other models. The conclusions are presented in Sec.~4.

\section{Non-perturbative input for valence quarks of pion and maximum entropy method with Tsallis entropy}
\label{sec:Entropy}
At the very low resolution scale, for $\pi ^ +$ meson, it is reasonable to consider that the dominant components are the up valence quark and the down valence anti-quark. As demonstrated in Ref.~\cite{Han:2018wsw}, at the initial scale $Q^2_0$, the widely used parameterized form of valence quarks can be expressed as
\begin{equation}
\label{Eq:parameterization}
u_{\nu}(x, Q_{0}^{2})=\bar{d}_{\nu}(x, Q_{0}^{2})=A_{\pi}x^{B_{\pi}}\left(1-x\right)^{C_{\pi}}.
\end{equation}
This parameterization can describe well both the Regge behavior at small $x$ and counting rule at large $x$. Here, $A_{\pi}$, $B_{\pi}$ and $C_\pi$ are undetermined parameters. In Eq.~(\ref{Eq:parameterization}) we assume that the isospin symmetry has not been broken and the mass difference between the $u_\nu$ and $\bar{d}_\nu$ is negligible, so the distributions of $u_\nu$ and $\bar{d}_\nu$ are identical~\cite{Han:2018wsw}.
In addition, the distributions in Eq.~(\ref{Eq:parameterization}) need to satisfy the valence sum rule
\begin{equation}
\label{Eq:quark sum rule}
\int\limits_{0}^{1}u_{\nu}(x,Q_{0}^{2})dx=\int\limits_{0}^{1}\bar{d}_{\nu}(x,Q_{0}^{2})dx=1.
\end{equation}
While the momentum sum rule is written as
\begin{equation}
\label{Eq:momentum sum rule}
\int\limits_{0}^{1}x[u_{\nu}(x, Q^{2}_0)+\bar{d}_{\nu}(x, Q^{2}_0)]dx=1-g,
\end{equation}
where the $g$ indicates the presence of gluons or sea quarks in the pion in addition to valence quarks. It is found that the $g$ is nonzero to obtain a better fit to the pion valence quark distribution as the sea quarks and the gluons can not be ignored even at some initial scale~\cite{Han:2018wsw}. The parameters $Q_0^2$ and $g$ exhibit a strong dependence on each other. At extremely low values of $Q_0^2$, dressed valence quarks can indeed be regarded as the sole effective degrees of freedom. As we move to relatively higher scales, the evolution process promptly generates distributions of sea quarks and gluons through quark splitting mechanisms. However, the determination of the initial scale value $Q_0^2$ remains contentious, as discussed in Ref. [21]. Consequently, we treat $g$ as a fitting parameter.

Now, with the constraints in Eqs.~(\ref{Eq:quark sum rule}) and ~(\ref{Eq:momentum sum rule}), the number of the undetermined parameters in Eq.~(\ref{Eq:parameterization}) reduced from three to one. In order to determine the remaining parameter, the maximum entropy method with Tsallis entropy is used. For the quark system, as defined in Eq.~(\ref{eq:tsallis_entropy}), the entropy for $u_\nu$ or $\bar{d}_\nu$ is
\begin{equation}
S_q(v)=\frac{1}{q-1}(1-\int(v(x))^q dx) , ~(q\neq1),
\end{equation}
where $v(x)$ is the valence quark distribution function and the $q$ is a free parameter. Hence, at the low resolution scale where valence quarks are predominant, the entropy of the $\pi^+$ system is defined as
\begin{equation}
\begin{array}{ c c l}
S_q^{\pi}&=&\frac{1}{q-1}(1-\int_{0}^{1}(u_{\nu}(x, Q^{2}))^{q}dx)+\frac{1}{q-1}(1-\int_{0}^{1}(\bar{d}_{\nu}(x, Q^{2}))^{q}dx)  \\
&&-\frac{1}{q-1}(1-\int_{0}^{1}(u_{\nu}(x, Q^{2}))^{q}dx)(1-\int_{0}^{1}(\bar{d}_{\nu}(x, Q^{2}))^{q}dx).  \\
\end{array}
\end{equation}

According to the maximum entropy principle, the optimal non-perturbative input in Eq.~(\ref{Eq:parameterization}) will have the largest entropy. Therefore, the remaining free parameter in Eq.~(\ref{Eq:parameterization}) will be determined by taking maximum entropy. However, $q$ is still an undetermined parameter derived from Tsallis entropy. To estimate the value of $q$, the valence quark distributions need to evolved to high $Q^2$ and compare with the experimental data. In the literature, the $Q^2$-dependence of PDFs are performed by the well-known Dokshitzer-Gribov-Lipatov-Altarelli-Parisi (DGLAP) evolution equation~\cite{Dokshitzer:1977sg,Altarelli:1977zs,Gribov:1971zn}. However, the linear DGLAP equation is valid in moderate $x$ and large $Q^2$. What is more, only parton splitting is considered in DGLAP equation, which give rise to the destruction of  unitary or the Froissart bound~\cite{Froissart:1961ux}. To address this issue, the parton recombination effect should be considered.  This effect is first considered by the Gribov-Levin-Ryskin-Mueller-Qiu (GLR-MQ) equation~\cite{Gribov:1983ivg,Mueller:1985wy} and its solution has been deeply investigated in recent years~\cite{Devee:2014fna,Cai:2024guq,Boroun:2023clh,Boroun:2023zvc}. Based on time-ordered perturbation theory (TOPT), Zhu, Ruan and Shen (ZRS) reconsider the parton recombination process~\cite{Zhu:1998hg,Zhu:1999ht,Zhu:2004xj} and obtain the modified DGLAP evolution equations with GLR-MQ-ZRS corrections, which satisfy the momentum conservation rule~\cite{Zhu:1999ht}. As the dominant effect is the recombination of gluon-gluon \cite{Wang:2016sfq}, in this study we utilize modified DGLAP equations with simplified GLR-MQ-ZRS corrections, expressed as
\begin{equation}
\label{eq:ZRS_valence}
Q^2\frac{dxf_{q_i}(x,Q^2)}{dQ^2}=\frac{\alpha_s(Q^2)}{2\pi}P_{qq}\otimes f_{q_i}
\end{equation}
for valence quark distributions,
\begin{equation}
\label{eq:ZRS_sea}
\begin{array}{ c c l}
Q^{2}\frac{dxf_{\bar{q}_{i}}(x,Q^{2})}{dQ^{2}}&=&\frac{\alpha_{s}(Q^{2})}{2\pi}[P_{qq}\otimes f_{\bar{q}_{i}}+P_{qg}\otimes f_{g}] \\
  &&-\frac{\alpha_{s}^{2}(Q^{2})}{4\pi R^{2}Q^{2}}\int_{x}^{1/2}\frac{dy}{y}xP_{gg\to\bar{q}}(x,y)[yf_{g}(y,Q^{2})]^{2}  \\
  &&+\frac{\alpha_{s}^{2}(Q^{2})}{4\pi R^{2}Q^{2}}\int_{x/2}^{x}\frac{dy}{y}xP_{gg\to\bar{q}}(x,y)[yf_{g}(y,Q^{2})]^{2}  \\
\end{array}
\end{equation}
for the sea quark distributions, and
\begin{equation}
\label{eq:ZRS_gluon}
\begin{array}{ c c l}
Q^{2}\frac{dxf_{g}(x,Q^{2})}{dQ^{2}}&=&\frac{\alpha_{s}(Q^{2})}{2\pi}[P_{gq}\otimes\Sigma+P_{gg}\otimes f_{g}]  \\
&&-\frac{\alpha_s^2(Q^2)}{4\pi R^2Q^2}\int_x^{1/2}\frac{dy}yxP_{gg\to g}(x,y)[yf_g(y,Q^2)]^2  \\
&&+\frac{\alpha_{s}^{2}(Q^{2})}{4\pi R^{2}Q^{2}}\int_{x/2}^{x}\frac{dy}{y}xP_{gg\to g}(x,y)[yf_{g}(y,Q^{2})]^{2} \\
\end{array}
\end{equation}
for the gluon distribution. Here, $R$ is the correlation radius of two gluons, $\Sigma$ is the singlet quark distribution, $P_{ij}$ is the splitting kernel functions \cite{Altarelli:1977zs,Wang:2016sfq}, and $P_{gg\to g}$, $P_{gg\to \bar{q}}$ are the gluon-gluon recombination kernels \cite{Wang:2016sfq}.

The running coupling $\alpha_{s}$ is a fundamental component in the formulation of evolution equations. In recent calculations of continuum Schwinger methods, a infrared-safe parametrization of saturating $\alpha_{s}$ was proposed \cite{Cui:2020tdf}. This saturating $\alpha_{s}$ was used to investigate the nucleon structure, which gives a unified description across the entire $Q^2$ range \cite{Wang:2024wny}. As our evolution proceeds from extremely low scale to a high scale,  we employed infrared-safe couplings to get rid of the divergent problem of $\alpha_{s}$ at low scale. This  saturating $\alpha_{s}$ is given by \cite{Cui:2020tdf}
\begin{equation}
\hat{\alpha}(k^{2})=\frac{\gamma_{m}\pi}{\ln\left[\frac{\mathcal{K}^{2}(k^{2})}{\Lambda_{\mathrm{QCD}}^{2}}\right]},~~~\mathcal{K}^{2}(y)=\frac{a_{0}^{2}+a_{1}y+y^{2}}{b_{0}+y},
\end{equation}
here $\gamma_{m}=4/[11-(2/3)n_{f}],~\Lambda_{\mathrm{QCD}}=0.234\mathrm{~GeV},~a_{0}=0.104\mathrm{~GeV}^{2},~a_{1}=0.0975\mathrm{~GeV}^{2},~b_{0}=0.121\mathrm{~GeV}^{2}$ and $n_{f}$ is the number of flavor. With the above evolution equations and saturating $\alpha_{s}$, we shall determine the pion PDFs.

\section{Results and discussions}
\label{sec:result}
In this section, we shall use the MEM with the Tsallis entropy to determine the valence quark distributions and use the modified DGLAP equations with the simplified GLR-MQ-ZRS corrections to perform the $Q^2$ evolution for comparing with the experimental data. As the determination of the initial scale value $Q^2_0$ remains a subject of debate, we treat it as a fitting parameter. Simultaneously, the parameter $g$ is designated as a free parameter to account for contributions from sea quarks and gluons at the initial scale value $Q^2_0$.

The Tsallis entropy for pion as the  function of $B_\pi$ and $q$ is shown in Fig.~1.  As shown in Fig.~1, it can be observed that $S_q^\pi$ monotonically decreases with q increasing, while entropy has a maximum in $B_\pi$ direction. Therefore, it is physically realizable to determine the valence quark distributions based on the MEM with the Tsallis entropy.

\begin{figure}[h]
\centering
\includegraphics[scale=0.5]{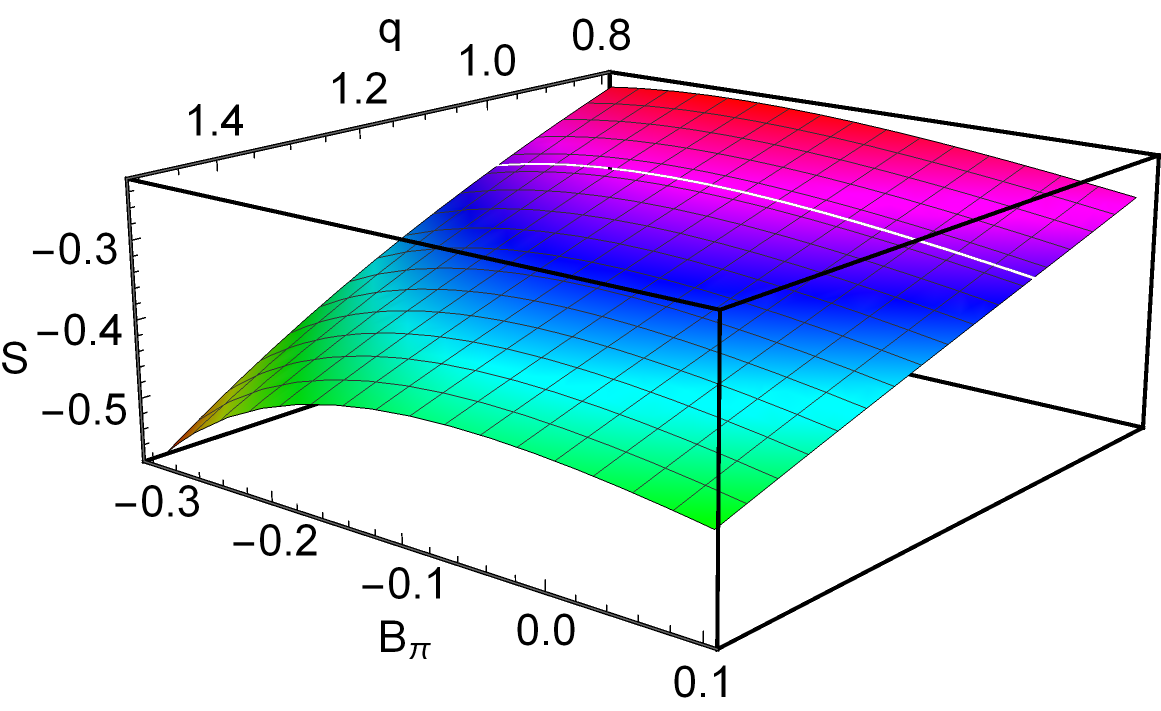}
\vspace*{-2mm}
\caption{The Tsallis entropy of valence quark nonperturbative input as functions of $B_\pi$ and $q$.}
\label{fig:tsallis_entropy}
\end{figure}

To more effectively illustrate the relationship between entropy and the undetermined parameters, we tentatively adopt a straightforward approach by fixing $g = 0.295$ ($Q^2_0=0.26\,\mathrm{GeV}^2$ ), consistent with the global fit presented in GRS99~\cite{Gluck:1999xe}. In Fig.~2, we present the results for Tsallis entropy as a function of $B_\pi$ at various values of $q$  at the initial scale. It is evident that the undetermined parameters are related to entropy, and these parameters can be determined through maximum entropy principles. However, since $q$ remains unknown, it must be obtained by fitting experimental data. Furthermore, the unfixed initial scale $Q^2_0$ exhibits a strong correlation with $g$; thus, we will treat $q$, $g$ and $Q^2_0$ as fitting parameters moving forward.

\begin{center}
\vspace{2mm}
\tabcolsep=10pt
\small
\renewcommand\arraystretch{1.2}
\begin{minipage}{\textwidth}
\centering
\textbf{Table 1.} Results from our MEM fits to the E615 data~\cite{Conway:1989fs} and the reanalyzed E615 data ~\cite{Aicher:2010cb}.
\end{minipage}
\vspace{5pt}
\begin{tabular}{|c|c|c|c|c|c|c|c|c|}
\hline
 & $Q^2_0$ & $g$ & $q$ & $A$ & $B$ & $C$ & $D$ & $\chi^2/\mathrm{d.o.f}$  \\
 \hline
Fit-A & $0.093$ & $0.0$ & $0.915$& $1.012$ & $0.006$ & $0.006$& $-$ & $0.680$ \\	
Fit-B & $0.096$ & $0.0$ & $0.917$& $1.153$ & $0.072$ & $0.121$& $0.151$ & $0.638$ \\
Fit-C & $0.142$ & $0.180$ & $0.928$& $ 0.724$ & $-0.132$ & $1.932$& $24.171$ & $1.445$ \\
\hline
\end{tabular}
\end{center}

\newpage

\begin{figure}[t]
\centering
\vspace*{-3mm}
\includegraphics[scale=1.0]{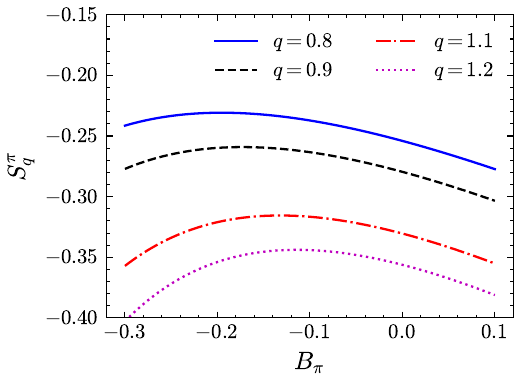}\hspace{4mm}%
\vspace*{-3mm}
\caption{The dependence of Tsallis entropy of valence quarks on $B_\pi$ at $q=0.8$ (blue solid curve), $q=0.9$ (black dashed curve), $q=1.1$ (red dot-dashed curve), $q=1.2$ (pink dotted curve) at initial scale.}
\label{fig:subfigures}
\end{figure}

In order to obtain the optimal value of parameters, the valence quark distribution functions at $Q^2=20~\mathrm{GeV}^2$ are calculated by modified DGLAP equation. The obtained $Q^2$ dependent valence quark distributions are compared with experimental data from E615~\cite{Conway:1989fs} (Fit-A). The parameters from Fit-A are listed in the first row of Table.~1. As shown in Table.~1, $Q_0^2 = 0.093 $ and $g = 0 $ yields a minimum value for $\chi^2$. This finding demonstrates that dressed valence quarks can indeed be regarded as effective degrees of freedom at extremely low-energy scales. The corresponding optimal value of $q$ is 0.915. This value of $q$ has a deviation from unity, which indicates that the correlation between valence quarks plays an important role in the investigation of pion valence quarks distribution.

\begin{figure}[th!]
\centering
\vspace*{-3mm}
\includegraphics[scale=0.4]{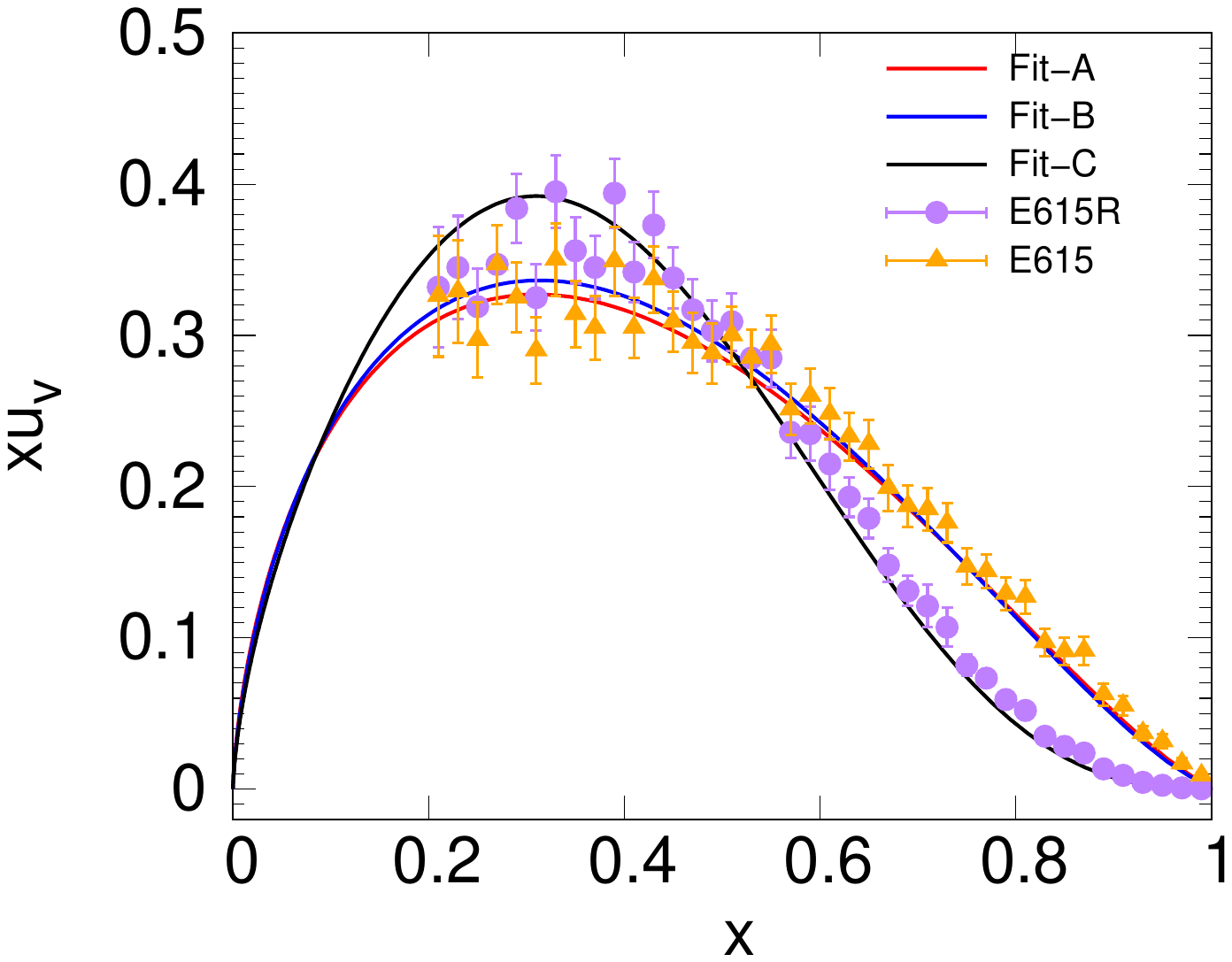}
\vspace*{-3mm}
\caption{The predicted up valence quark distribution function is presented in comparison with the E615 data~\cite{Conway:1989fs} and E615R data~\cite{Aicher:2010cb} at a scale of $Q^2=20\,\mathrm{GeV^2}$.}
\label{fig:e615}
\end{figure}

In Fig.~3, we present the up valence quark distribution function at $Q^2=20\,\mathrm{GeV^2}$ by performing the modified DGLAP evolution with the initial distribution in Eq.~(\ref{Eq:parameterization}). From the red line in Fig.~3, one can see that the valence quark distribution function from Fit-A shows a good agreement with the E615 data~\cite{Conway:1989fs}. It should be noted that the E615 data were reanalyzed in Ref.~\cite{Aicher:2010cb}, resulting in a significantly softer valence distribution at large-$x$. Utilizing this modified dataset (E615R), we conducted a fitting using Eq.~(\ref{Eq:parameterization}). However, we found that the fitting was inadequate. This may be attributed to the simplicity of the functional form, which requires an additional factor to suppress the valence quark distribution. Consequently, we refit the experimental data employing the same valence quark distribution form as utilized in Ref.~\cite{Aicher:2010cb}.
\begin{equation}
\label{Eq:modified-parameterization}
u_{\nu}(x, Q_{0}^{2})=\bar{d}_{\nu}(x, Q_{0}^{2})=A_{\pi}x^{B_{\pi}}\left(1-x\right)^{C_{\pi}}(1+Dx^2).
\end{equation}

Using Eq.~(\ref{Eq:modified-parameterization}), we conducted fittings for the data from E615 (Fit-B) and E615R (Fit-C). As demonstrated by the blue and black lines in Fig. 3, Fit-B and Fit-C effectively characterized the E615he and E615R data, respectively. From the last two rows of Table.~1, we can see that, compared with the fit to E615, the fit to E615R yields a higher $g$ value, which makes the softer quark distribution at initial  $Q^2$. It is important to note that when employing various distribution functions from different models to describe the $J/\psi$ production data, it has been observed that the $J/\psi$ data tend to favor a higher gluon distribution~\cite {Chang:2020rdy}. Table.~1 demonstrates an interesting result, that is, in the three sets of fitting results, the $q$ values all deviate significantly from unity, which indicates that the consideration of correlation between valence quarks is reasonable.

To test this outcome, we calculate the moments of valence quark distributions, which have been well calculated from LQCD and some other models~\cite{deTeramond:2018ecg,Cui:2020tdf,Gao:2020ito,Han:2018wsw,Davidson:1994uv,Sutton:1991ay}. The moments of the momentum fraction of valence quark distribution is defined as
\begin{equation}
<x^n>=\int\limits_{0}^{1}x^nu_{\nu}(x, Q^2)dx,
\end{equation}
where the $n$ is the order of the moment. The lowest three nontrivial moments of valence quark distribution at $Q^2=4~\mathrm{GeV}^2$ are listed in Table.~2. As shown in Table.~2, our predictions are consistent with the results from different models.

\vspace{2mm}
\begin{center}
\tabcolsep=15pt
\small
\renewcommand\arraystretch{1.2}
\begin{minipage}{15.5cm}{
\small{\bf Table 2.} The list of the first three moments of valence quark momentum distributions at $Q^2=4~\mathrm{GeV}^2$ in comparision with those from QCD analyses~\cite{Sutton:1991ay}, DSE~\cite{Cui:2020tdf}, LQCD~\cite{Gao:2020ito}, LFHQCD~\cite{deTeramond:2018ecg}, NJL model~\cite{Davidson:1994uv} and Shannon~\cite{Han:2018wsw}.}
\end{minipage}
\vglue5pt
\begin{tabular}{| c | c | c | c | }
\hline
 & $\left\langle x \right\rangle$ & $\left\langle x^2 \right\rangle$ & $\left\langle x^3 \right\rangle$  \\
 \hline
QCD analyses~\cite{Sutton:1991ay} & $0.23$ & $0.099$ & $0.055$\\	
DSE~\cite{Cui:2020tdf} & $0.24$ & $0.094$ & $0.047$ \\
LQCD~\cite{Gao:2020ito} & $0.213$ & $0.101$ & $0.0061$ \\
LFHQCD~\cite{deTeramond:2018ecg} & $0.233$ & $0.103$ & $0.056$ \\
NJL model~\cite{Davidson:1994uv}& $0.236$ & $0.103$ & $0.057$ \\
Von Neumann entropy~\cite{Han:2018wsw} & $0.24$ & $0.10$ & $0.057$ \\
This work, Fit-A & $0.238$ & $0.105$ & $0.058$ \\
This work, Fit-B & $0.242$ & $0.106$ & $0.059$ \\
This work, Fit-C & $0.233$ & $0.091$ & $0.044$ \\
\hline
\end{tabular}
\end{center}

\section{Summary}
\label{sec:summary}

In this study, we investigate two distinct functional forms of initial non-perturbative input for valence quark distributions at a low resolution scale \(Q^2_0\). The parameters are determined using two well-established constraints: the valence sum rule and the momentum sum rule, along with the MEM employing Tsallis entropy. By comparing the \(Q^2\)-dependent valence quark distributions derived from the modified DGLAP equation, which includes GLR-MQ-ZRS corrections, to experimental data, we find that our results exhibit good agreement with both E615 and E615R datasets when utilizing the input specified in Eq.~(\ref{Eq:modified-parameterization}). Additionally, the moments of the valence quark distributions are calculated and found to be consistent with results from different models at $Q^2=4\,\mathrm{GeV}^2$.

In the field of information theory, entropy is utilized as a metric for quantifying uncertainty. The concept of entropy is applicable and useful in determining the internal structure of hadron since the detailed information of the valence quark distribution is unknown. By imposing constraints from well-known properties, the MEM aims to maximize entropy in order to ensure unbiased inference of the probability distribution. In MEM, the Tsallis entropy demonstrates a powerful ability in determining pion valence quark distributions. The value of $q$ deviates from unity, which indicates that the valence quarks have correlation even at low $Q^2$.

\vspace*{2mm}

\section*{Acknowledgments}
This work is supported by National Key R$\&$D Program of China (Grant NO. 2024YFE0109800 and 2024YFE0109802); National Natural Science Foundation of China under Grant No. 12375073; Guizhou Provincial Basic Research Program (Natural Science) under Grant No. QKHJC-ZK[2023]YB027.

\end{document}